\documentclass{JHEP3} 
\usepackage{epsfig}
\epsfclipon

\def\ssubsubsection#1{\vspace{3mm} \noindent \textbf{#1} \\ \vspace{-3mm} \\ \noindent}

\newcommand{\nco}{\newcommand}
\nco{\beq}{\begin{equation}} \nco{\eeq}{\end{equation}}
\nco{\beqa}{\begin{eqnarray}} \nco{\eeqa}{\end{eqnarray}}
\nco{\lra}{\leftrightarrow}
\def\sfrac#1#2{{\textstyle{#1\over #2}}}

\nco{\sss}{\scriptscriptstyle} \nco{\dphi}{\varphi}
\nco{\lsim}{\mbox{\raisebox{-.6ex}{~$\stackrel{<}{\sim}$~}}}
\nco{\gsim}{\mbox{\raisebox{-.6ex}{~$\stackrel{>}{\sim}$~}}}

\def\pref#1{(\ref{#1})}

\def\Hp{H_+}
\def\hp{h_+}
\def\Hm{H_-}
\def\hm{h_-}

\def\kbr{{\overline{k}}}

\title{\Large Effective Field Theories and Inflation}

\author{C.P.\ Burgess, J.M. Cline\\
Physics Department, McGill University,
3600 University Street, Montr\'eal, Qu\'ebec, Canada H3A 2T8\\
E-mail: \email{cliff@physics.mcgill.ca},
\email{jcline@physics.mcgill.ca}, }
\author{and R.~ Holman\\ Physics Department, Carnegie Mellon
University, Pittsburgh PA 15213\\
E-mail: \email{rh4a@andrew.cmu.edu}}

\preprint{McGill-03/12}


\abstract{We investigate the possible influence of
very-high-energy physics on inflationary predictions focussing on
whether effective field theories can allow effects which are
parametrically larger than order $H^2/M^2$, where $M$ is the scale
of heavy physics and $H$ is the Hubble scale at horizon exit. By
investigating supersymmetric hybrid inflation models, we show that
decoupling does not preclude heavy-physics having effects for the
CMB with observable size even if $H^2/M^2 \ll O(1\%)$, although
their presence can only be inferred from observations given some
{\it a priori} assumptions about the inflationary mechanism. Our
analysis differs from the results of hep-th/0210233, in which
other kinds of heavy-physics effects were found which could alter
inflationary predictions for CMB fluctuations, inasmuch as the
heavy-physics \emph{can} be integrated out here to produce an
effective field theory description of low-energy physics. We
argue, as in hep-th/0210233, that the potential presence of
heavy-physics effects in the CMB does {\it not} alter the
predictions of inflation for generic models, but {\it does} make
the search for deviations from standard predictions worthwhile. }

\begin{document}

\section{Introduction and Discussion \label{section:intro}}
The recent inflationary literature contains considerable
discussion of whether or not detailed observations of fluctuations
in the temperature of the Cosmic Microwave Background (CMB) can be
used to infer the properties of extremely high energies ---
usually assumed to be above the Planck scale. This discussion is
particularly timely due to the arrival of ever-more-accurate
measurements of these fluctuations \cite{boomerang,dasi}, most
recently by the Wilkinson Microwave Anisotropy Probe (WMAP)
collaboration \cite{WMAP}. There is likely to be even further
improvement in the not-too-distant future \cite{Planck}.

Broadly speaking, two points of view have emerged from this
discussion.
\begin{enumerate}
\item CMB fluctuations can depend on the details of high-energy
(trans-Planckian) physics, and so represents an opportunity to
probe these otherwise inaccessible scales \cite{tp1,tp2,tp3}.
\item General decoupling arguments require the influence of physics
at scale $M \gg H$ to contribute at most of order $H^2/M^2$ to
observable late-time effects, where $H$ is the Hubble scale at
horizon exit. This precludes the intrusion of higher-energy
physics into CMB fluctuations \cite{shenker}.
\end{enumerate}

Clearly much is at stake. On the one hand, given specific guesses
for what trans-Planckian physics might be, observable effects for
the CMB have been calculated. Although some of these calculations
remain controversial \cite{dSinconsistent} -- in particular the
choice of nonstandard vacua in de Sitter space -- it is hard to
argue trans-Planckian physics cannot interfere with inflationary
predictions without better understanding what trans-Planckian
physics is.

On the other hand, if general decoupling arguments do not apply in
an inflationary context then the very {\it predictability} of
inflationary models is lost. Indeed, since the nature of the
higher-energy physics is at present unknown any of its
implications must be included under the general heading of
theoretical uncertainty when comparing with experiments. In this
sense decoupling is a prerequisite for using the properties of the
CMB as evidence for an earlier inflationary period
\cite{WMAPinflation} in the first place.

To the extent that the central issue is the validity and
implications of decoupling during inflation, it may be addressed
without invoking unknown trans-Planckian physics. The conclusions
drawn might then also be applicable to trans-Planckian physics to
the extent that it also satisfies the assumptions of the analysis.
Here (and in \cite{us}) we use simple sub-Planckian field theories
to explore these issues. Within this context two separate
questions may be addressed.
\begin{enumerate}
\item Is it required that higher-energy physics decouple at all,
in the sense that its low-energy effects must be described in
terms of a low-energy effective field theory?
\item Given that the low-energy implications of heavy physics
{\it can} be described by an effective field theory during the
epoch of horizon exit, need its influence for the CMB be limited
to effects which are of order $H^2/M^2$?
\end{enumerate}
Ref.~\cite{us} addresses the first of these questions, and shows
that oscillating background fields before horizon exit can
invalidate an effective-field-theory description. They can do so
by preventing the time evolution of the relevant modes of the
inflaton from being adiabatic, which is also an obstacle to using
an effective-lagrangian description outside of a cosmological
context.

In this paper we address the second question of whether an
effective field theory can produce detectable effects even if
those of order $H^2/M^2$ are too small to be observable. We here
show that for some models heavy physics generates contributions to
the inflaton potential which depend logarithmically on the heavy
mass $M$, and so contribute to the slow-roll parameters an amount
of order $M_0^2/M^2$, with $M_0 \gg H$. In so doing we also show
that since the CMB measurements in themselves only probe the
inflaton potential in a limited way, at present the influence of
heavy physics can only be inferred using CMB measurements given
some sort of {\it a priori} assumptions about the nature of the
physics which is responsible for making the inflaton potential
flat in the first place.

Although the thrust of both of these conclusions is that
high-energy physics {\it can} intrude into CMB fluctuations, we
argue that they do so in a way which does {\it not} introduce
uncontrollable theoretical errors into the predictions of
inflationary models, and so do not undermine the successful
comparison of these predictions with observations. They do not do
so because our basic conclusion is that the criteria for
decoupling in inflation are precisely the same criteria which
apply in other non-cosmological contexts. In particular, the vast
majority of high-energy effects {\it do} decouple and so cannot
alter standard predictions. But just as in other areas of physics,
some specific kinds of effective interactions can be sensitive to
higher-energy details and these are worth scrutinizing for the
information they may contain about the microscopic physics of very
small distances. In this sense our results, like those of
\cite{us}, represent in some ways the best of all possible worlds.

Our presentation is organized as follows. Section \ref{sec:toy}
starts with a toy model of hybrid inflation consisting of two
scalar fields having very different masses. We use this model to
explicitly integrate out the heavier of the two (the heavy
physics) at one loop to see its effects on the lighter scalar (the
inflaton). This example illustrates first that, in the generic
case heavy-field loop contributions to the inflaton potential are
large and dangerous, since they tend to ruin inflation by
destroying the flatness of the inflaton potential. This is one of
the well-known naturalness problems of inflation, and any careful
treatment of the effects for inflation of higher-energy physics
must properly address this.

We can also use this example to show that even if the above
naturalness problems are addressed, there is a further obstruction
to identifying heavy-physics effects within the observed
fluctuations of the CMB if the inflaton is really rolling slowly
at the epoch of horizon exit. This is because the inflationary
predictions in this case only sample the first few derivatives of
the inflaton potential, and these can generically be adjusted by
small changes in the renormalizable inflaton couplings. Because of
this a detection of heavy-physics effects in inflation is only
possible within the context of a specific inflationary model, for
which the inflaton couplings can be subject to {\it a priori}
conditions (such as those due to symmetries).

Section \ref{sec:flat} sharpens the analysis by addressing these
two issues within a supersymmetric model. Supersymmetry provides
an attractive framework for addressing the above questions because
supersymmetry both controls the naturalness issue and provides
{\it a priori} constraints on the form of the low-energy inflaton
potential. Section \ref{subsec:global} examines these issues
within a standard globally supersymmetric hybrid-inflation model.
The light scalar for this model parameterizes a precisely flat
direction of the classical potential, whose degeneracy is lifted
purely by the virtual effects of heavy fields. In this model it is
therefore purely heavy-physics effects which make the difference
between the success and failure of the theory as a model of
inflation, since it is their small size which explains why the
inflaton potential is very shallow (but not exactly flat). Simple
modifications of this higher-energy physics produce sizeable
changes in the predictions for the CMB precisely because in these
models it is the higher-energy physics which controls the entire
effect.

Section \ref{subsec:sugra} generalizes the model of section
\ref{subsec:global} to supergravity. The main message of this
section is that such an extension is possible despite the
well-known $\eta$ problem of supergravity models. We illustrate
this using the example of D-term inflation.

We emphasize that all of the examples we use are extremely
orthodox, and have been considered in various contexts in the
literature. Our purpose in bringing them all together here is to
first show that existing models already provide explicit examples of how
higher-energy physics can have important effects during inflation,
and so to illustrate that such effects can happen in ordinary
theories which decouple, without relying on other exotic
properties (like violations of Lorentz invariance, for example).
A secondary goal is to raise the bar for putative models of
trans-Planckian physics, by arguing that any serious candidate for
this must: ($i$) establish that the proposed trans-Planckian
physics does decouple from lower-energy phenomena so as to not
completely sacrifice the predictivity of lower-energy physics, and
($ii$) contain a precise model of the low-energy inflaton dynamics
as a benchmark against which the high-energy physics effects can
be compared.

\section{Inflation in a Toy Effective Theory}\label{sec:toy}
In this section we accomplish two things using a simple toy model
involving an inflaton $\phi$ and some heavy fields $\chi_i$.
First, we explicitly integrate out the heavy fields to derive the
dominant terms in the resulting low-energy effective field theory
for $\phi$. Then we apply this result to an inflationary model and
show how these heavy-field contributions are poison to
inflationary models, since they destroy the flatness of the
inflaton potential. This well-known naturalness problem shows that
the virtual effects of high-energy fields for inflation are
generically too {\it large} rather than too small. This
calculation is standard \cite{RobertRMP} and is mainly used to
motivate the calculations of the next section, where we repeat the
analysis using supersymmetric examples. A reader familiar with
these issues, and who is in a hurry, can skip directly to section
\ref{sec:flat}.

\subsection{The Model}\label{subsec:model}
Consider the following toy model of very-high-energy physics,
which is a very minor generalization of a standard
hybrid-inflation model \cite{hybrid}:
\beqa \label{eq:modeldef}
    - \, {{\cal L} \over \sqrt{-g}} &=&  \sfrac12 \, \partial_\mu  \phi \,
    \partial^\mu \phi + \sfrac12 \, \partial_\mu  \chi_i \, \partial^\mu \chi_i +
    V(\phi,\chi)  , \\
    \hbox{with} \qquad V(\phi,\chi) &=& V_{\rm inf}(\phi) - \sfrac12 \,
    m^2 \chi_i^2 + \sfrac12 \, g \, \chi_i^2 \phi^2 . \nonumber
\eeqa
We assume the $N$ massive scalars,\footnote{We introduce $N$ heavy
fields simply to amplify the effects of the $\chi$ fields, and
much of our discussion applies equally well if $N=1$. Perturbative
quartic self-interactions for $\chi$ can also be included without
materially affecting our discussion.} $\chi_i, i=1,...,N$
--- which represent the heavy physics whose lower-energy influence
we wish to determine --- satisfy $\langle \chi_i \rangle = 0$,
thereby excluding the non-adiabatic effects discussed in
ref.~\cite{us}. For the simplicity of later formulae we choose the
couplings to be invariant under the $O(N)$ which rotates the
$\chi_i$'s amongst themselves. In eq.~\pref{eq:modeldef} $\phi$ is
the inflaton, whose potential
\beq \label{eq:Vinf}
    V_{\rm inf}(\phi) = \rho + \sfrac12 \, m_\phi^2 \, \phi^2 +
    \sfrac14 \, \lambda \, \phi^4
\eeq
is chosen to ensure sufficient $e$-foldings of expansion before an
eventual inflationary exit and reheating. This is ensured if the
two slow-roll parameters \cite{liddlelyth} satisfy\footnote{We use
here the rationalized Planck mass, $M_p^{-2} = 8 \pi G$.}
\beqa \label{eq:slowroll}
    \epsilon_0(\phi) &=& \frac12 \, \left( {M_p \, V'_{\rm inf} \over
    V_{\rm inf}} \right)^2 = \left( {M_p \, \phi \, (m_\phi^2 +
    \lambda \, \phi^2) \over \rho + \sfrac12 \, m_\phi^2 \, \phi^2 +
    \sfrac14 \, \lambda \, \phi^4 } \right)^2 \ll 1 \nonumber \\
    \eta_0(\phi) &=& \left( {M_p^2 \, V_{\rm inf}'' \over
    V_{\rm inf}} \right) = \left( {M_p^2 \, (m_\phi^2 + 3
     \lambda \, \phi^2) \over \rho + \sfrac12 \, m_\phi^2 \, \phi^2 +
    \sfrac14 \, \lambda \, \phi^4 } \right) \ll 1 \, .
\eeqa
For instance, these conditions are satisfied if $m_\phi^2 M_p^2
\ll \rho$ and if we choose initial conditions for which $\chi_i =
0$ and $\rho/m_\phi^2 \gg \phi^2 \gg m^2/g$ (provided $\lambda$ is
small enough also to ensure $\lambda \phi^2 \ll m_\phi^2$
throughout this range of $\phi$). Inflation then occurs with $H^2
= V_{\rm inf}(\phi)/(3 M_p^2) \approx \rho/(3 M_p^2)$ while $\phi$
rolls slowly down to $\phi_{\rm end}$, where inflation ends.
Inflation ends either when the slow-roll parameters become of
order unity, or when $\phi^2 \sim m^2/g$, at which point $\chi_i$
moves quickly away from zero, and $\phi_{\rm end}$ is the field
determined by whichever of these occurs first.

In order to study the effects of high energy physics we assume the
masses,
\beq
    \label{eq:Mphi}
    M^2(\phi) = -m^2 + g \, \phi^2,
\eeq
to be much larger than the Hubble scale, $H^2$, during horizon
exit. In order to keep the analysis under control we also assume
the $\chi_i$ fields to be sub-Planckian throughout the
inflationary slow roll of $\phi$: $M^2(\phi) \ll M_p^2$.

\subsection{Integrating out the Heavy Fields}\label{subsec:heavy}
Under the above assumptions the heavy fields may be explicitly
integrated out and an effective-field-theory analysis applies.
Since our main interest is in effects which do not decouple we
focus on those terms in the effective theory which are
unsuppressed by powers of $1/M$. These come in two types: ($i$)
relevant interactions, which are proportional to positive powers
of $M$; or ($ii$) marginal interactions, which grow
logarithmically with $M$.

A simple calculation of the virtual effects of the heavy scalars
on the light scalar potential is obtained by matching the one-loop
corrected effective potential for the full theory with the
one-loop effective potential in the theory involving only $\phi$,
but including effective interactions. This gives the following
result:
\beqa \label{eq:effpot}
    V_{\rm eff}(\phi) &=& V_{\rm inf}(\phi) + \Delta V(\phi), \\
    \hbox{with} \qquad \Delta V(\phi) &=& V_{\rm ct}(\phi) + \frac{N}{64 \pi^2}
    M^4(\phi) \, \ln \left( {M^2(\phi) \over \mu^2} \right) \, ,
    \nonumber
\eeqa
where $V_{\rm inf}$ now denotes the renormalized inflaton
potential, eq.~\pref{eq:Vinf}, with constants $\rho(\mu)$,
$m_\phi^2(\mu)$ and $\lambda(\mu)$ chosen according to a
renormalization prescription which is described in detail below.
$V_{\rm ct}$ contains the counter-terms, $\delta \rho$, $\delta
m_\phi^2$ and $\delta \lambda$, which enforce the renormalization
condition, and $\mu$ is a floating scale which also depends on the
renormalization scheme. As usual, the implicit dependence of
$V_{\rm inf}$ on $\mu$ is just such as to ensure that $\mu$
cancels in physical observables.

For inflationary purposes it is convenient to cast the
renormalization conditions in terms of the slow-roll parameters
since we must really demand that $V_{\rm eff}$, rather then
$V_{\rm inf}$, produces sufficient inflation. For our laterpurposes, a convenient way to do so is to require that $V_{\rm
eff}$ and $V_{\rm inf}$ share the same value for the Hubble
constant and the two slow-roll parameters at a particular point in
field space, which we can choose to be the point when a specific
mode $k_*$ leaves the horizon. We denote the time of horizon exit
for this mode by $t_*$, where $H(t_*) = k_{*{\rm phys}} =
k_*/a(t_*)$, and we further denote the value of the inflaton field
at this time by $\phi(t_*) = \phi_*$. Then this renormalization
condition states that $V_{\rm eff}(\phi_*)$, $\epsilon(\phi_*)$
and $\eta(\phi_*)$ are given by their tree-level expressions in
terms of $\rho$, $m_\phi$ and $\lambda$:
\beqa \label{eq:rencond}
    V_* &=& V_{\rm eff}(\phi_*) = \rho + \sfrac12 \, m_\phi \, \phi_*^2 +
    \sfrac14 \, \lambda \, \phi_*^4 = 3 \, M_p^2 \, H_*^2\nonumber \\
    \epsilon_* &=& \epsilon_0(\phi_*) = \frac12 \, \left[ {M_p \, \phi_* \,
    ( m_\phi^2 + \lambda \, \phi_*^2) \over V_{\rm eff}(\phi_*)} \right]^2
    = \frac12 \, \left[ {\phi_* \,( m_\phi^2
    + \lambda \, \phi_*^2) \over 3 \, M_p \, H_*^2} \right]^2   \\
    \eta_* &=& \eta_0(\phi_*) = {M_p^2 \, (m_\phi^2 + 3 \, \lambda \, \phi_*^2)
    \over V_{\rm eff}(\phi_*)} = {(m_\phi^2 + 3 \, \lambda \, \phi_*^2)
    \over 3 \, H_*^2} \, . \nonumber
\eeqa
The subscript `0' on the quantities $\epsilon_0(\phi)$ and
$\eta_0(\phi)$ indicates that their functional dependence is as
calculated using the potential $V_{\rm inf}$, as in
eq.~\pref{eq:slowroll}.

These conditions amount to the following requirements for $\Delta
V$: $\Delta V(\phi_*) = \Delta V'(\phi_*) = \Delta V''(\phi_*) =
0$, and so
\beq \label{eq:DeltaV}
    \Delta V(\phi) = {N \over 64 \pi^2} \left\{ M^4(\phi) \left[
    \ln \left( {M^2(\phi) \over M^2_*} \right) - \sfrac32 \right]
    + \sfrac12 \, M_*^2 \Bigl[4 \, M^2(\phi) - M_*^2 \Bigr]
    \right\},
\eeq
where $M_* = M(\phi_*)$.

Using the following derivatives:
\beqa \label{eq:dVeff}
    \Delta V'(\phi) &=& {g \, N \, \phi \over 16 \pi^2}
    \left[ (m^2 + g \, \phi^2) \ln \left( {M^2 \over M_*^2} \right) -
    g ( \phi^2 - \phi_*^2 ) \right] ,\nonumber \\
    \Delta V''(\phi) &=& { g \, N \over 16 \pi^2 } \left[ (m^2 + 3 g \,
    \phi^2) \ln \left( {M^2 \over M_*^2} \right) - g( \phi^2 -
    \phi_*^2) \right] , \nonumber \\
    \Delta V'''(\phi) &=& {g^2 \, N \, \phi \over 8 \pi^2} \left[
    3 \ln \left( { M^2 \over M_*^2} \right) + {2 g \,
    \phi^2 \over M^2} \right],
\eeqa
and taking $M^2 \approx g \phi^2 \gg m^2$ and $\Delta V \ll V_{\rm
inf} \approx \rho$, we find $H^2 \approx \rho/(3 M_p^2)$ and the
following expressions for the slow-roll parameters:
\beqa \label{eq:dslowroll}
    (2 \, \epsilon)^{1/2} &\approx& (2 \, \epsilon_0)^{1/2} + {g^2 \, N \, \phi \over
    48 \pi^2 \, M_p \, H^2} \left[ \phi^2 \ln \left( {\phi^2 \over \phi_*^2} \right) -
    ( \phi^2 - \phi_*^2 ) \right] \nonumber \\
    \eta &\approx& \eta_0 + { g^2 \, N  \over 48 \pi^2 \, H^2} \left[ 3\,
    \phi^2 \ln \left( {\phi^2 \over \phi_*^2} \right) - ( \phi^2 -
    \phi_*^2) \right] \, ,
\eeqa
as well as the second-order slow-roll quantity:
\beq \label{eq:xidef}
    \xi^2  = (2 \, \epsilon)^{1/2} \, \left( {M_p^3 V''' \over V} \right) \approx
    (2 \, \epsilon)^{1/2} \, {M_p \, \phi \over H^2} \,
    \left\{ 2 \lambda + \, {g^2 \, N  \over 24 \pi^2 } \left[
    3 \ln \left( { \phi^2 \over \phi_*^2} \right) + 2 \right] \right\}.
\eeq

For our purposes, there are two lessons to be learned from these
expressions: the naturalness problems they imply, and the
obstruction they raise to the inference of heavy-physics
properties purely using measurements of the CMB.

\ssubsubsection{Naturalness Problems}\label{subsubsec:naturalness}
The biggest difficulty with these expressions is in maintaining
inflation itself. Although we've ensured (by construction) that
the $\phi$ motion is sufficiently slow near $\phi = \phi_*$, this
is not in itself sufficient to obtain the more than 50
$e$-foldings of inflation which are required for successful
cosmology. In particular, this requires $\eta$ to remain small
over a significant range of $\phi$, which requires
\beq \label{eq:dphivsH}
   \left| \phi^2 - \phi^2_* \right| \ll {48 \pi^2  \,
   H^2 \over g^2 \, N} \, .
\eeq
This is typically difficult to satisfy unless $g$ is extremely
small. For instance, if we take $H$ to be approximately constant
over the $N_e$ $e$-foldings of inflation between horizon exit and
inflation's end, and if we take $dV_{\rm inf}/d\phi \approx
m_\phi^2 \, \phi$ over this region, we have $\phi_*^2 \approx
(m^2/g) \exp\left[(2 m_\phi^2 \, N_e)/(3 H^2) \right]$. In this
case we see eq.~\pref{eq:dphivsH} implies
\beq
    \left| \exp\left[ \frac{2 m_\phi^2 \, N_e}{3 H^2} \right] - 1
    \right| \ll \frac{48 \pi^2}{g \, N} \,
    \left( \frac{H^2}{m^2} \right) \, ,
\eeq
and so even if we have already assumed $H^2/m_\phi^2 \sim N_e
\gsim 50$, we must now in addition tune $m$ to satisfy $N m^2 \ll
48 \pi^2 H^2/g$.

We see that heavy loops can have big effects because they compete
with unusually small low-energy interactions. The low-energy
interactions are small precisely because inflation requires the
low-energy inflaton potential must be chosen to be so very flat.
In these circumstance generic kinds of heavy physics not only do
not decouple, they can completely ruin inflation.

\ssubsubsection{Detecting Heavy Physics Using the CMB}\label{subsubsec:detect}
Eq.~\pref{eq:dslowroll} implies another obstruction to learning
about higher-energy physics using inflationary predictions for the
CMB, independent of the naturalness issues associated with
obtaining the slow roll itself. This obstruction arises because if
the inflaton describes a sufficiently slow roll, its effects for
the CMB are completely determined by the quantities $H$,
$\epsilon$ and $\eta$ evaluated at horizon exit. For instance
standard expressions \cite{liddlelyth} for the corrections to the
CMB fluctuation spectrum are\footnote{Note that there is
 a sign error in the formula for $\xi$ in
ref.~\cite{liddlelyth}}
\beqa \label{eq:spectrum}
    \delta_H^2(k_*) &\approx& {V_* \over 150 \pi^2 \, M_p^4 \,
    \epsilon_*}
    = {H^2_* \over 50 \pi^2 \, M_p^2 \, \epsilon_*}
    \nonumber \\
    n(k_*) &\approx& 1 + 2 \, \eta_* - 6 \, \epsilon_* \\
    {d n \over d \ln k}(k_*) &\approx& 16 \, \epsilon_* \, \eta_* - 24 \,
    \epsilon^2_* - 2 \, \xi^2_* . \nonumber
\eeqa
Since the above renormalization scheme is designed not to change
the slow-roll parameters at horizon exit, it ensures that the
heavy fields cannot alter the inflationary predictions for modes
near $k = k_*$, to leading order in the slow-roll parameters. To
the extent that this is true, and that observations are only
sensitive to fluctuation properties near $k_*$, any detection of
heavy physics as a distortion of the CMB spectrum must rely on the
breakdown of the slow-roll conditions near horizon exit (such as
might be true if the preliminary WMAP indications for nonzero
$dn/d\ln k$ \cite{WMAPnrun} should prove to be significant).

Until more detailed observational information is available, we
conclude that the effects of very heavy physics for the CMB can at
present only be inferred relative to some {\it a priori}
information about the nature of inflationary physics. If, for
instance, inflation were believed to be due to a particular
mechanism --- such as perhaps the string-motivated brane-inflation
proposals of \cite{BI} --- then detailed knowledge of the form for
$V_{\rm inf}$ can allow sufficiently large corrections to this
form to be inferred from observations.

An extreme example of {\it a priori} constraints is the case where
$V_{\rm inf}$ is precisely constant, in which case the inflaton
potential is {\it entirely} due to loop-generated heavy-physics
effects. This situation is actually fairly common for
supersymmetric models, which are typically rife with
classically-flat directions. In these models the inflaton can be a
modulus parameterizing one of these directions, and so the very
flatness of the inflaton potential is then partially explained by
its origin as a loop-generated effect.

\section{Flat Directions and Supersymmetric Models}\label{sec:flat}
In this section we repeat the calculation of the previous section
for a supersymmetric example, for which the generic naturalness
issues raised above can be controlled. This allows us to more
precisely compute the size of high-energy loop effects on CMB
fluctuations, and so to illustrate how these effects need not be
unobservably small even if their scale is much higher than $H$.
They also provide examples wherein the entire inflaton potential
arises as such a heavy-physics loop effect, and so for which the
very detection of inflationary effects in the CMB is necessarily
also a detection of heavy-physics effects. We first describe
models in global supersymmetry in section \ref{subsec:global}, for
which the calculations are simple. We shall find self-consistency
forces us then to generalize these to supergravity, and this is
done in section \ref{subsec:sugra}.

The results we obtain in this section are simple to state. In the
models we consider here, the tree-level inflaton potential is
exactly flat, but this flat direction is lifted by virtual loops
of heavy particles yielding nonzero slow-roll parameters $\eta$
and $\epsilon$. In particular, integrating out the heavy particles
produces slow-roll parameters which are suppressed by factors of
order $M_0^2/M^2$, where $M$ is the relevant heavy mass scale, and
it is this decoupling which makes the slow-roll parameters small.
But, and this is the crux of the matter, the reference scale $M_0$
is much bigger than $H$, and so these models may be taken as an
existence proof that heavy physics can decouple and yet still
alter inflationary predictions for the CMB since the figure of
merit for deciding the observability of the heavy-physics effects
can be larger than $H^2/M^2$.

\subsection{Globally Supersymmetric Models}\label{subsec:global}
Part of the attraction of supersymmetric models for hybrid
inflation is the ubiquity with which their scalar potentials have
flat directions. Two kinds of models have been proposed, which
differ according to whether the inflationary potential arises as
an $F$-term \cite{Fterm} or a $D$-term \cite{Dterm}. We focus here
on $D$-term models in order to avoid the usual $\eta$ problem when
we generalize to supergravity.

Consider, then, a model containing the chiral multiplets, $\Phi =
\{\phi,\psi\}$ and $H_\pm = \{h_\pm,\chi_\pm\}$, coupled to a
$U(1)$ gauge multiplet, $V = \{A_\mu,\lambda\}$. We take $\Hp$ and
$\Hm$ to carry opposite $U(1)$ charge $\pm e$, and the multiplet
$\Phi$ to be neutral. The model's superpotential and K\"ahler
potential are
\beq
    K = \Hp^* \Hp + \Hm^* \Hm + \Phi^* \Phi \qquad \hbox{and}
    \qquad W = g \Phi \, (\Hp \, \Hm - v^2) \, ,
\eeq
where $g$ and $v$ are real constants. The associated scalar
potential is $V = V_F + V_D$ where
\beqa
    V_F &=& g^2 \left( \Bigl| \hp \hm - v^2 \Bigr|^2 + \Bigl|  \phi \,
    \hm \Bigr|^2 + \Bigl|  \phi \,  \hp \Bigr|^2 \right) \, , \nonumber \\
    V_D &=& \frac{e^2}{2} \, \Bigl( |\hp|^2 - |\hm|^2 + \xi
    \Bigr)^2 \, ,
\eeqa
where $\xi > 0$ is the Fayet-Iliopoulos term. Notice that $V_D =
0$ implies $|\hm|^2 = |\hp|^2 + \xi$ for any $\phi$ while $V_F =
0$ implies $\phi = 0$ and $\hp \hm = v^2$, so the global minimum
is supersymmetric and has
\beq
    \phi = 0, \qquad |h_\pm|^2 = \frac12 \Bigl( \mp \xi + \sqrt{\xi^2
    + 4 v^4} \Bigr) \, .
\eeq

The feature of most interest for the present purposes is the
potential's trough at $h_\pm = 0$ and large $|\phi|$, for which
$V_{\rm trough}(\phi) = V(h_\pm = 0, \phi) = g^2 \, v^4 + \sfrac12
\, e^2 \xi^2$. For sufficiently large $|\phi|$ this is a local
minimum in the $h_\pm$ directions, with scalar excitations in
these directions having masses
\beq
    M_\pm^2(\phi) = g^2 |\phi|^2 \pm \sqrt{ g^4 v^4 + e^4 \xi^2} \, .
\eeq
Everywhere along this trough the scalar $\phi$ is precisely
massless, while linear combinations of the complex scalars $h_\pm$
and $h^*_\mp$ are massive, with masses, $M_\pm^2(\phi)$, which are
large for large $|\phi|$. Since the $U(1)$ gauge invariance is not
broken along the trough, the gauge bosons are massless. The
fermion masses along the trough, on the other hand, are zero for
the gaugino, $\lambda$, and the chiral-multiplet fermion $\psi$.
They are nonzero for the fermions $\chi_\pm$, with mass
eigenvalues
\beq
    m^2_\pm(\phi) = g^2 |\phi|^2 \, .
\eeq

Along the trough's bottom the tree-level spectrum therefore breaks
up into a sector of massless particles,
$\{A_\mu,\lambda,\phi,\psi\}$, which do not classically directly
couple among themselves, but which do couple to a massive sector,
$\{\hp,\chi_+,\hm,\chi_-\}$. In this section we integrate out this
massive sector to determine the effective interactions which are
generated in this way amongst the light fields, with the goal of
using this as a candidate hybrid inflation model \cite{gShybrid}.

At the classical level the model does {\it not} describe viable
hybrid inflation, with $\phi$ interpreted as the inflaton rolling
along the trough's bottom. This is because at the classical level
the $\phi$ potential is precisely flat, and so there is nothing to
drive the inflaton's slow roll. This is no longer true once  loop
corrections are included since the virtual heavy fields will
generate an inflaton potential.

Following the same procedure as for the previous section we obtain
the heavy-field contribution to the inflaton potential by matching
the one-loop results. It is useful to extend the model to include
$N$ charged chiral multiplets, $H^i_\pm$, in an $O(N)$-invariant
way, in which case we find the result
\beqa \label{eq:effpotgs}
    V_{\rm eff}(\phi) &=& \rho + \Delta V(\phi), \\
    \hbox{with} \qquad \Delta V(\phi) &=& \delta \rho +\frac{2N}{64
    \pi^2} \sum_{i=\pm} \left[ M_i^4(\phi) \, \ln \left( {M_i^2(\phi)
    \over \mu^2} \right) - m_i^4(\phi) \, \ln \left( {m_i^2(\phi)
    \over \mu^2} \right) \right] \, ,
    \nonumber
\eeqa
where $\rho$ is the renormalized (constant) classical potential
along the trough (which classically is $\rho = g^2 \, v^4 +
\sfrac12 \, e^2 \xi^2$), and $\delta \rho$ is the corresponding
counter-term.

It is convenient to evaluate this result using the previous
expressions $M^2_\pm(\phi) = m^2(\phi) \pm \Delta$ and
$m^2_\pm(\phi) = m^2(\phi)$, where $m(\phi) = g|\phi|$ and $\Delta
= \sqrt{g^4 v^4 + e^4 \xi^2}$. In terms of these the potential
becomes
\beqa \label{eq:effpotgsexpl}
    \Delta V_{\rm eff}(\phi) = \delta \rho &+& \frac{N}{32
    \pi^2} \left[m^4(\phi) \, \ln \left( {m^4(\phi) - \Delta^2
    \over m^4(\phi)} \right) + \Delta^2 \, \ln \left( {m^4(\phi)
    - \Delta^2 \over \mu^4} \right) \right. \nonumber\\ &+&
    \left. 2 m^2(\phi) \Delta \, \ln \left(
    {m^2(\phi) + \Delta
    \over m^2(\phi) - \Delta} \right) \right] \, ,
    \nonumber
\eeqa
which for $m^2(\phi) \gg \Delta$ becomes
\beq
    \Delta V_{\rm eff}(\phi) \approx  \frac{N \, \Delta^2}{16
    \pi^2}\left[ \ln \left( {m^2(\phi) \over m_*^2} \right)
    + {\cal O} \left( {\Delta^2 \over m^4 } \right) \right] \, ,
\eeq
where $m_*^2 = m^2(\phi_*) = g^2 |\phi_*|^2$ and we adopt the
renormalization condition that $\Delta V$ must vanish when $\phi =
\phi_*$.

This induced potential causes the inflaton $\varphi = |\phi|$ to
roll, and this roll can be slow if $|\phi|$ is sufficiently large,
since
\beqa
\label{sr}
    (2 \epsilon)^{1/2} &=& {M_p \over V_{\rm
    eff}} \left( {\partial V_{\rm eff} \over
    \partial \varphi} \right) \approx {M_p \,N \, \Delta^2
    \over 8 \pi^2 \rho \, \varphi}
     \, , \nonumber \\
    \eta &=& {M_p^2 \over V_{\rm eff}} \left( {\partial^2
    V_{\rm eff} \over \partial \varphi^2} \right)
    \approx - \, {M_p^2 \,N\, \Delta^2 \over 8 \pi^2
    \rho\, \varphi^2}
    \, .
\eeqa

With these expressions we can examine more quantitatively the
conditions for inflaton. For simplicity we take for these purposes
$e \sim g$ and $\xi \sim v^2$, which implies $\rho \sim g^2 v^4$
and $\Delta \sim g^2 v^2$ and so $\Delta^2/\rho \sim g^2$. We also
choose $N=1$, although we return to other choices for $N$ at the
end. Suppose now $N_e$ $e$-foldings of inflation occur between
horizon exit and the end of inflation. During this time the field
$\phi$ evolves from $\phi_*$ to $\phi_{\rm end}$, with
\beq
    \phi_*^2 \approx \phi^2_{\rm end} + \frac{N N_e
    \Delta^2}{12 \pi^2 H^2} \approx \phi^2_{\rm end} +
    \frac{g^2 N N_e \, M_p^2}{4 \pi^2} ,
\eeq
where $\phi_{\rm end}^2$ is defined by the condition that the
fields $h_\pm$ start to roll ($\phi^2_{\rm end} \sim \Delta/g^2$)
or that the slow-roll parameter $\eta$ becomes order unity
($\phi^2_{\rm end} \sim g^2 N M_p^2/(8 \pi^2)$, whichever is
reached first. It turns out that $\Delta/g^2$ is reached first if
$g \gsim 10^{-3}$, and it is to this case we now specialize for
concreteness' sake.

With the above choices, and taking $\phi_{\rm end}^2 \ll
\phi_*^2$, the amplitude of scalar fluctuations becomes
\beq
    \delta_s^2 = \frac{H^2}{50 \pi^2 M_p^2 \, \epsilon_*} \sim
    \left( \frac{16 N_e \, v^4}{75 N M_p^4} \right)\, ,
\eeq
which for 60 $e$-foldings agrees with the CMB value, $\delta_s
\sim 10^{-5}$ if $v/M_p \sim g \sim 10^{-3}$. The slow-roll
parameters at horizon exit similarly become
\beq
    \epsilon_* \approx \frac{g^2 N}{32 \pi^2 N_e} \, ,
    \qquad\qquad
    \eta_* \approx -\; \frac{1}{2 N_e} \, .
\eeq
As is easily checked, $\phi_*$ (and so also $m_*$) is less than
$M_p$ provided $g^2 N N_e/(4 \pi^2) < O(1)$. This is important
because so long as $\phi \ll M_p$ we are justified to restrict our
analysis to global supersymmetry, instead of supergravity.

In this model $\epsilon_* \ll \eta_*$, so $n_s-1 \approx 2\eta
\sim 10^{-2}$ (if $N_e \sim 50$). This is perfectly compatible
with the WMAP data; for example with the choice of cosmological
parameters $\Omega_b = 0.046$, $\Omega_{cdm} = 0.178$,
$\Omega_{\Lambda} = 1-\Omega_b-\Omega_{cdm}$, $h=0.72$, $\tau=
0.17$ and $n_s=0.99$ in CMBFAST \cite{cmbfast}, we obtain $\chi^2
= 1430$ using the WMAP likelihood code \cite{verde}. This is as
good a fit to the data as the WMAP preferred parameter values
\cite{map-params}. Moreover since $\epsilon\ll\eta$ the model is
not constrained by gravitational wave production, with the WMAP
gravity-wave constraint giving $16\epsilon = r < 1.3$
\cite{map-inf}, implying the weak constraint $\epsilon < 0.08$. It
is interesting to note that other simple models of inflation, such
as $\lambda\phi^4$ chaotic inflation, are close to being ruled out
by the CMB data due to having too large a tilt
\cite{map-inf,others}.

Having established that the model is experimentally viable, let us
return to the issue of decoupling. The implications of decoupling
are most easily seen by writing the slow-roll parameters as
\beq
    (2 \epsilon_*)^{1/2} \approx \frac{g^2 N \Delta^2}{8 \pi^2 \, \rho} \, \left(
    \frac{\varphi_* \, M_p}{m^2_*} \right) \qquad \hbox{and}
    \qquad \eta_*  \approx -\,\frac{g^2 N \Delta^2}{8 \pi^2 \,
    \rho} \, \left(\frac{M_p^2}{m^2_*} \right) \,,
\eeq
which shows that both are suppressed by two powers of the heavy
mass $m_*$. What is important is that the scale against which
$m^2_*$ is compared is not $H^2$, but is instead $(g^2 N
\Delta^2/8\pi^2 \rho) M_p^2 \sim (g^4 N/8 \pi^2) M_p^2$. As we
dial the parameters of the model so that $m_*$ is becomes larger,
the slow roll parameters get closer to the scale-invariant point
$\epsilon_* = \eta_* = 0$. This is a consequence of decoupling,
since the entire inflaton potential for the model is generated by
virtual effects of the heavy physics. But because the benchmark
for observability is {\it not} $H^2/m^2_*$, the difference from
scale invariance can be kept observable even if $H^2/m_*^2$ is
much smaller than a few percent.

In particular, we can consider two models which differ only in
their spectrum of particles at scale $m_*$, by comparing the
effects two theories which differ only in the number of heavy
multiplets. We take $N = 1$ for model 1 and $N=2$ for model 2, and
we suppose both models to undergo the same number of $e$-foldings
of inflation, $N_e$. With these choices these models agree on
their predictions for $\eta_*$. They also predict identical
fluctuation amplitudes provided $v_1^4 = v_2^4/2$, and so if they
further share the same couplings, $g_1 \sim g_2$, then we have
$\epsilon_{*2} = 2 \epsilon_{*1}$. Clearly the two models
therefore can have detectable differences in their predictions for
CMB observables, provided only that $\epsilon_*$ is larger than
$O(0.01)$.

\subsection{Supergravity Models}\label{subsec:sugra}
This section makes a technical point by showing that a model
similar to the previous hybrid inflation model can be embedded
into supergravity without losing its main features. We do so
because we have seen that inflationary applications can (but need
not, for $g \ll 1$) require fields $|\phi| \gg M_p$, for which
supergravity corrections to the action are generally not
negligible.

The ability to make the extension from global to local
supersymmetry is not trivial, and would have been much more
difficult if we had adopted an example relying purely on the
scalar potential's $F$-terms. In the $F$-term case the
supergravity corrections typically introduce new terms which are
of order $H^2 |\phi|^2$, and so which contribute an ${\cal O}(1)$
amount to the slow-roll parameter $\eta$ --- a result known as the
supersymmetric $\eta$-problem \cite{etaproblem}. Our main purpose
here is to show how this problem is evaded in the case of $D$-term
inflation \cite{Dterm}.

Recall for these purposes the form of the supergravity scalar
sector \cite{sugra}. Given the K\"ahler function, $K(z,z^*)$, and
superpotential, $W(z)$, the kinetic and potential terms for a
collection of chiral scalars, $z^i$, are:
\beq
    {\cal L} = - \sqrt{-g} \left[ \frac12 \, K_{i}^{j^*}(z,z^*) \,
    \partial_\mu z^i \partial^\mu z^*_j + V(z,z^*) \right] \, ,
\eeq
and $V = V_F + V_D$ with
\beq
    V_F = e^{K/M_p^2} \, \left[ \hat{K}^i_{j^*} (D_i W) (D^j W)^* -
    \frac{3 \, |W|^2}{M_p^2} \right] \, ,
\eeq
and
\beq
    V_D = \frac12 \, f^{ab} \, \Bigl[ K_i (T_a z)^i + \xi_a \Bigr]
    \, \Bigl[ K_j (T_b z)^j + \xi_b \Bigr] \, .
\eeq
Here $K_i^{j^*} = \partial^2 K/\partial z^i \partial z^*_j$ and
$\hat{K}^i_{j^*}$ is its inverse matrix. Similarly $K_i = \partial
K/\partial z^i$, $T_b$ is a gauge-group generator, $\xi_a$ is a
Fayet-Iliopoulos term (and so is only present for $U(1)$
gauge-group factors) and $f^{ab}$ is the inverse matrix of Re
$F_{ab}(z)$, where $F_{ab}$ is the holomorphic gauge-kinetic
function. In these expressions the K\"ahler derivative is
\beq
    D_i W = \frac{\partial W}{\partial z^i} + \frac{W}{M_p^2}
    \frac{\partial K}{\partial z^i}  = W_i + \frac{K_i \, W}{M_p^2} \,
    .
\eeq

We now specialize as before to an inflaton multiplet, $\Phi$, plus
two electrically charged multiplets, $H_\pm$. Write $\varphi =
(\Phi + \Phi^*)/M_p$ and take the no-scale ansatz \cite{noscale}
\beq
    K(\varphi,\Hp^* \Hp, \Hm^* \Hm)  = - 3 \, M_p^2 \, \log\varphi +
     k(\Hp^* \Hp) + \kbr(\Hm^*\Hm) \, ,
\eeq
$F_{ab} = \delta_{ab}$ and $W = W(\Hp,\Hm)$. With these choices we
have $W_\Phi = \partial W/\partial \Phi = 0$ and the K\"ahler
derivatives are:
\beqa
    D_\Phi W &=& - \, \frac{3 \, W}{M_p \, \varphi} \nonumber\\
    D_{\Hp} W &=& W_{\Hp} + {k' \Hp^* \, W \over M_p^2} \, , \\
    D_{\Hm} W &=& W_{\Hm} + {\kbr' \Hm^* \, W \over M_p^2} \, ,
\eeqa
where $k' = \partial k/\partial x$, for $x = \Hp^* \Hp$, and so
on. For instance, if $k = \Hp^* \Hp$ then $k' = 1$ {\it etc.}

The K\"ahler metric is
\beq
    K_i^{j^*} = \pmatrix{ 3/\varphi^2 & 0 & 0 \cr 0 & k' + k''
    \Hp^*\Hp
    & 0 \cr 0 & 0 & \kbr' + \kbr'' \Hm^* \Hm \cr} \,
\eeq
and so its inverse matrix becomes
\beq
    \hat{K}^i_{j^*} =
    \pmatrix{\varphi^2/3 & 0 & 0 \cr 0 & 1/(k' + k'' \Hp^*\Hp) & 0 \cr
    0 & 0 & 1/(\kbr' + \kbr'' \Hm^* \Hm )\cr } \, .
\eeq
Positivity of the kinetic energies requires $k' + k'' \Hp^*\Hp >
0$ and $\kbr' + \kbr'' \Hm^* \Hm > 0$.

With these choices the scalar potential, $V = V_F + V_D$, becomes
\beq
    V_F = \frac{e^{(k + \kbr)/M_p^2}}{\varphi^3} \,
    \left[ {|D_{\Hp} W|^2 \over k' + k'' \Hp^*\Hp}
    + {|D_{\Hm} W|^2 \over \kbr' + \kbr'' \Hm^* \Hm} \right] \,
    ,
\eeq
and
\beq
    V_D = \frac{e^2}{2} \left[ k'|\Hp|^2 - \kbr' |\Hm|^2 + \xi \right]^2\,,
\eeq
where we take a $U(1)$ gauge group with Fayet-Iliopoulos term
$\xi$.

For inflationary purposes, we may further specialize to
\beq
    W = M \, (\Hp \Hm - v^2) \, ,
\eeq
and so $W_{\Hp} = M \Hm$ and $W_{\Hm} = M \Hp$. Also suppose both
$k'$ and $\kbr'$ are finite and nonzero as $H_\pm \to 0$, and that
both $k''$ and $\kbr''$ are also finite (but possibly zero) in
this limit. In this case we have $D_{\Hp} W \to 0$ and $D_{\Hm} W
\to 0$ as $H_\pm \to 0$, and so also $V_F \to 0$ in this limit.

Any zero of both $V_F$ and $V_D$ is a global minimum for $V$,
which in our case is obtained by choosing $k' |\Hp|^2 + \xi =
\kbr' |\Hm|^2$ to ensure $V_D = 0$, and then choosing $\varphi \to
\infty$ to make $V_F = 0$. (This solution exists, for instance,
for the minimal case $k' = \kbr' = 1$.) In this limit
supersymmetry is broken (since $D_iW \ne 0$) but with $V = 0$
classically in the limit of large $\varphi$.

There is also a trough with nonzero $V$ which is independent of
$\phi$, corresponding to the case $H_\pm = 0$. In this limit we
have $D_{H_\pm} W = 0$, and so $V_F = 0$, and so $V = V_D = e^2
\xi^2/2$. The mass of the $H_\pm$ scalar fields about this trough
are both positive and of order $M$, provided that $v \ll M_p$ and
$M \gg e \, \sqrt\xi$. This trough is the direct analogue of the
trough considered above in the globally-supersymmetric case, and
we may apply a similar analysis again here to the same effect.
Just as for the globally-supersymmetric case we have a classically
flat potential, whose degeneracy gets lifted by virtual loops of
heavy particles, and so for which the relative effect of a heavy
multiplet for the slow-roll parameters need not be small.

The effective potential for this SUGRA model is similar to that of
the previous SUSY model, with an important qualitative difference.
Here the field-dependent scalar field masses for $H_\pm$ take the
form
\beq
    M^2\pm = {M^2\over \phi^3}\mp e^2\xi
\eeq
while their superpartners have masses $m^2_\pm = {M^2\over
\phi^3}$.  In the limit where ${M^2\over \phi^3}\gg  e^2\xi$, the
effective potential is approximately
\beq
    V_{\rm eff} = \frac12{e^2}\xi^2 \left(1 - {3e^2\over
    8\pi^2}\ln{\phi\over\phi_+}\right)
\eeq
Therefore $\phi$ rolls to larger instead of smaller values, which
is expected since the global minimum is at $\phi\to\infty$ in this
model.  Unlike the SUSY model, the slow roll approximation only
gets better as the inflaton rolls, so the end of inflation is
definitely triggered by the instability of $H_+$, at a field value
given by
\beq
    \phi_e^3 = {M^2\over e^2\xi}
\eeq
The smallest allowable initial value of $\phi$ is determined by
the observational constraint on $n_s-1 \cong 2\eta\lsim 0.2$,
\beq
    \phi_0^2 = {3e^2\over 8\pi^2 \eta}
\eeq
and the relation between $\phi$ and the number of e-foldings,
$\phi^2 = \phi_0^2 + {3e^2\over 4\pi^2} N_e$,  gives the
constraint that
\beq
\label{e5}
    \phi_e^2 - \phi_0^2 = {3e^2\over 4\pi^2} N_e \quad\to \quad
    e^5 = {M^2\over \xi}
    \left( {1\over 4\pi^2}\left(N_e + {3\over 2\eta}\right)\right)^{-3/2}
\eeq
The COBE normalization implies
\beq
    \xi = 5\times 10^{-4}{\sqrt{3\eta}\over 4\pi} M_p^2
\eeq
independently of the coupling $e^2$; thus $\xi$ is safely below
the Planck scale.  In fact, this model has the possiblity of
choosing $M$ and $\phi_e$ to be subPlanckian if desired, since
$M^2$ is freely adjustable.  If we choose $M$ such that $\phi_e
<1$, then \pref{e5} implies that $e^3 <
(4\pi^2/(N_e+3/2\eta))^{3/2}$, which is a weak constraint on the
coupling.  Even though the global minimum is at the superPlanckian
value $\phi\to\infty$, all the inflationary dynamics can be
comfortably accomplished in the regime where $\phi < 1$.

\section{Conclusions}\label{sec:conclusions}

What we hope to have made clear is that some care must be taken
when trying to use decoupling to limit the influence of heavy
physics on low energy observables. We have argued that for a class
of inflationary models, the inflaton potential is \emph{solely}
due to the effects of integrating out the heavy particles of the
model. This allows for changes in the CMB parameters of order
unity, instead of the $H^2\slash M^2$ effects that might na\"ively
be expected. Together with the non-adiabatic mechanism of
ref.~\cite{us} this gives two ways for heavy physics to be
relevant to the physics of horizon exit, using only garden-variety
hybrid inflation models.

Although our calculations are purely sub-Planckian, our
conclusions do allow some inferences concerning trans-Planckian
physics. If the trans-Planckian physics decouples, it may be able
to take advantage of the same kinds of mechanisms we have found to
leave some residue of itself in the CMB temperature anisotropies.
If, on the other hand, the trans-Planckian physics does not
decouple (as the $\alpha$-vacua proposals may not do) then its
low-energy implications are likely to be everywhere, and the first
problem is really to understand why low-energy predictions have
been possible at all to this point, before worrying specifically
about its implications for the CMB.

\acknowledgments

We would like to acknowledge the Aspen Center for Physics where
this work was begun. We thank Robert Brandenberger, Brian Greene,
Nemanja Kaloper, Anupam Mazumdar and Steve Shenker for helpful
discussions. R.~H. was supported in part by DOE grant
DE-FG03-91-ER40682, while C.B. and J.C. partially supported by
grants from McGill University, N.S.E.R.C. (Canada) and F.C.A.R.
(Qu\'ebec).


\begin{thebibliography}{99}

\bibitem{boomerang}http://www.physics.ucsb.edu/~boomerang/

\bibitem{dasi}http://astro.uchicago.edu/dasi/

\bibitem{WMAP}
http://map.gsfc.nasa.gov

\bibitem{Planck} http://astro.estec.esa.nl/Planck/

\bibitem{tp1}
J.~Martin and R.~H.~Brandenberger, ``The trans-Planckian problem
of inflationary cosmology,'' Phys.\ Rev.\ D {\bf 63}, 123501
(2001) [arXiv:hep-th/0005209];
R.~H.~Brandenberger and J.~Martin, ``The robustness of inflation
to changes in super-Planck-scale physics,'' Mod.\ Phys.\ Lett.\ A
{\bf 16}, 999 (2001) [arXiv:astro-ph/0005432];
%
R.~H.~Brandenberger and J.~Martin, ``On the Dependence of the
Spectra of Fluctuations in Inflationary Cosmology on
Trans-Planckian Physics'' [arXiv:hep-th/0305161].

\bibitem{tp2}
R.~Easther, B.~R.~Greene, W.~H.~Kinney and G.~Shiu, ``Inflation as
a probe of short distance physics,'' Phys.\ Rev.\ D {\bf 64},
103502 (2001) [arXiv:hep-th/0104102];
R.~Easther, B.~R.~Greene, W.~H.~Kinney and G.~Shiu, ``Imprints of
short distance physics on inflationary cosmology,''
[arXiv:hep-th/0110226];
R.~Easther, B.~R.~Greene, W.~H.~Kinney and G.~Shiu, ``A generic
estimate of trans-Planckian modifications to the primordial  power
spectrum in inflation,'' Phys.\ Rev.\ D {\bf 66}, 023518 (2002)
[arXiv:hep-th/0204129];

\bibitem{tp3}
C.~S.~Chu, B.~R.~Greene and G.~Shiu, ``Remarks on inflation and
noncommutative geometry,'' Mod.\ Phys.\ Lett.\ A {\bf 16}, 2231
(2001) [arXiv:hep-th/0011241];
J.~Martin and R.~H.~Brandenberger, ``A cosmological window on
trans-Planckian physics,'' [arXiv:astro-ph/0012031];
T.~Tanaka, ``A comment on trans-Planckian physics in inflationary
universe,'' [arXiv:astro-ph/0012431];
J.~C.~Niemeyer and R.~Parentani, ``Trans-Planckian dispersion and
scale-invariance of inflationary  perturbations,'' Phys.\ Rev.\ D
{\bf 64}, 101301 (2001) [arXiv:astro-ph/0101451];
A.~Kempf and J.~C.~Niemeyer, ``Perturbation spectrum in inflation
with cutoff,'' Phys.\ Rev.\ D {\bf 64}, 103501 (2001)
[arXiv:astro-ph/0103225];
A.~A.~Starobinsky, ``Robustness of the inflationary perturbation
spectrum to trans-Planckian  physics,'' Pisma Zh.\ Eksp.\ Teor.\
Fiz.\  {\bf 73}, 415 (2001) [JETP Lett.\  {\bf 73}, 371 (2001)]
[arXiv:astro-ph/0104043];
M.~Bastero-Gil, ``What can we learn by probing trans-Planckian
physics,'' [arXiv:hep-ph/0106133];
M.~Lemoine, M.~Lubo, J.~Martin and J.~P.~Uzan, ``The stress-energy
tensor for trans-Planckian cosmology,'' Phys.\ Rev.\ D {\bf 65},
023510 (2002) [arXiv:hep-th/0109128];
R.~H.~Brandenberger, S.~E.~Joras and J.~Martin, ``Trans-Planckian
physics and the spectrum of fluctuations in a bouncing universe,''
[arXiv:hep-th/0112122];
J.~Martin and R.~H.~Brandenberger, ``The Corley-Jacobson
dispersion relation and trans-Planckian inflation,'' Phys.\ Rev.\
D {\bf 65}, 103514 (2002) [arXiv:hep-th/0201189];
U.~H.~Danielsson, ``A note on inflation and trans-Planckian
physics,'' Phys.\ Rev.\ D {\bf 66}, 023511 (2002)
[arXiv:hep-th/0203198];
S.~F.~Hassan and M.~S.~Sloth, ``Trans-Planckian effects in
inflationary cosmology and the modified  uncertainty principle,''
[arXiv:hep-th/0204110];
U.~H.~Danielsson, ``Inflation, holography and the choice of vacuum
in de Sitter space,'' JHEP {\bf 0207}, 040 (2002)
[arXiv:hep-th/0205227];
A.~A.~Starobinsky and I.~I.~Tkachev,
JETP Lett.\  {\bf 76}, 235 (2002) [Pisma Zh.\ Eksp.\ Teor.\ Fiz.\
{\bf 76}, 291 (2002)] [arXiv:astro-ph/0207572];
K.~Goldstein and D.~A.~Lowe, ``Initial state effects on the cosmic
microwave background and  trans-planckian physics,''
[arXiv:hep-th/0208167];
U.~H.~Danielsson, ``On the consistency of de Sitter vacua,''
[arXiv:hep-th/0210058];
G.~Shiu and I.~Wasserman, ``On the signature of short distance
scale in the cosmic microwave  background,'' Phys.\ Lett.\ B {\bf
536}, 1 (2002) [arXiv:hep-th/0203113];

\bibitem{shenker}
N.~Kaloper, M.~Kleban, A.~Lawrence,
S.~Shenker, ``Signatures of short distance physics in the cosmic
microwave background'', [arXiv:0201158];
%
 N.~Kaloper, M.~Kleban,
A.~Lawrence, S.~Shenker, and L.~Susskind,
 ``Initial conditions for inflation'', [arXiv:hep-th/0209231].

\bibitem{dSinconsistent}
T. Banks and L. Mannelli, ``De Sitter vacua, renormalization and
locality'', Phys.\ Rev.\ {\bf D67} {2003} {065009}, [arXiv:hep-th/0209113];
%
M. Einhorn and F. Larsen, ``Interacting quantum field theory in de
Sitter vacua'', Phys. \ Rev. \ {\bf D67} {2003} {024001}, [arXiv:hep-th/0209159];
%
M. Einhorn and F. Larsen, ``Squeezed States in the de Sitter Vacuum'', [arXiv: hep-th/0305056];
%
K.~Goldstein and D.~A.~Lowe, ``A note on alpha-vacua and
interacting field theory in de Sitter space'' ,[arXiv:
hep-th/0302050].
%
Hael Collins, R.~Holman and Matthew~R.~Martin, ``The Fate of the
$\alpha$-Vacuum'', [arXiv:hep-th/0306028].

\bibitem{WMAPinflation}
H.V. Peiris, {\it et. al.}, [arXiv:astro-ph/0302225];
%
V. Barger, H.-S. Lee and D. Marfatia, [arXiv:hep-ph/0302150];
%
B. Kyae and Q. Shafi, [arXiv:astro-ph/0302504];
%
J.R. Ellis, M. Raidal and T. Yanagida, [arXiv:hep-ph/0303242].

\bibitem{us}
C.P. Burgess, J.M. Cline, F. Lemieux and R. Holman, JHEP 0302
(2003) 048 (24 pages) [ArXiv: hep-th/0210233].

\bibitem{RobertRMP}
R.H.~Brandenberger, Rev.\ Mod.\ Phys.\ {\bf 57} (1985) 1.

\bibitem{hybrid}
A.~Linde, Phys.\ Rev. {\bf D49} (1994) 748
[arXiv:astro-ph/9307002];
%
David H.~ Lyth and Ewan Stewart,`` More varieties of hybrid
inflation'', Phys.\ Rev.\ {\bf D54} (1996) {7186},
[arXiv:hep-ph/9606412];

\bibitem{liddlelyth}
Andrew R.~ Liddle and David H.~ Lyth
``Cosmological Inflation and Large Scale Structure'', Cambridge
University Press (2000).

\bibitem{WMAPnrun}
D.N. Spergel {\it et.al.}, [arXiv:astro-ph/0302209].

\bibitem{BI}
G.R. Dvali and S.H.Henry Tye, Phys.\ Lett.\ {\bf B450} (1999)
72-82 [arXiv:hep-ph/9812483];
%
A. Lukas, B.A. Ovrut and D. Waldram, Phys.\ Rev.\ {\bf D61} (2000)
023506 [arXiv:hep-th/9902071];
%
J.M. Cline, Phys.\ Rev.\ {\bf D61} (2000) 023513
[arXiv:hep-ph/9904495];
%
E.E. Flanagan, S.H.Henry Tye and I. Wasserman, Phys.\ Rev.\ {\bf
D62} (2000) 024011 [arXiv:hep-ph/9909373];
%
A. Kehagias and E. Kiritsis, JHEP (1999) 9911:022
[arXiv:hep-th/9910174];
%
J. Khoury, B.A. Ovrut, P.J. Steinhardt and N. Turok, Phys.\ Rev.\
{\bf D64} (2001) 123522 [arXiv:hep-th/0103239];
%
C.P. Burgess, M. Majumdar, D. Nolte, F. Quevedo and G. Rajesh,
JHEP 0107 (2001) 047 [arXiv:hep-th/0105204];
%
C. Herdeiro, S. Hirano and R. Kallosh, JHEP 0112 (2001) 027
[arXiv:hep-th/0110271];
%
C.P. Burgess, P. Martineau, F. Quevedo, G. Rajesh and R.J. Zhang,
JHEP 0203 (2002) 052 [arXiv:hep-th/0111025];
%
J. Garcia-Bellido, R. Rabadan and F. Zamora, JHEP 0201 (2002) 036
[arXiv:hep-th/0112147];
%
R. Blumenhagen, B. Kors, D. Lust and T. Ott, Nucl.\ Phys.\ {\bf
B641} (2002) 235-255 [arXiv:hep-th/0202124];
%
K. Dasgupta, C. Herdeiro, S. Hirano and R. Kallosh, Phys.\ Rev.\
{\bf D65} (2002) 126002 [arXiv:hep-th/0203019];
%
J.-Y. Kim, Phys.\ Lett.\ {\bf B548} (2002) 1-8
[arXiv:hep-th/0203084];
%
N. Jones, H. Stoica and S.H.Henry Tye, JHEP 0207 (2002) 051
[arXiv:hep-th/0203163].


\bibitem{lythriotto} David H.~ Lyth and Antonio Riotto, ``Particle Physics
Models of Inflation and the Cosmological Density Perturbation'',
\prep{314}{1}{1999}.

\bibitem{Fterm}
M. Bastero-Gil and S.F. King, Nucl.\ Phys.\ {\bf B549} (1999)
391-406 [arXiv:hep-ph/9806477];
%
J.A. Casas, G.B. Gelmini and A. Riotto, Phys.\ Lett.\ {\bf B459}
(1999) 91-96 [arXiv:hep-ph/9903492];
%
S.M. Harun-or-Rashid, T. Kobayashi and H. Shimabukuro, Phys.\
Lett.\ {\bf B466} (1999) 95-99 [arXiv:hep-ph/9908266];
%
J. McDonald, JHEP 0212 (2002) 029 [arXiv:hep-ph/0201016].

\bibitem{Dterm}
P. Binetruy and G.R. Dvali, Phys.\ Lett.\ {\bf B388} (1996)
241-246, [arXiv:hep-ph/9606342];
%
E. Halyo, Phys.\ Lett.\ {\bf B387} (1996) 43-47
[arXiv:hep-ph/9606423];
%
T. Matsuda, Phys.\ Lett.\ {\bf B423} (1998) 35-39
[arXiv:hep-ph/9705448];
%
D.H. Lyth and A. Riotto, Phys.\ Lett.\ {\bf B412} (1997) 28-34
[arXiv:hep-ph/9707273];
%
C.F. Kolda and J. March-Russell, Phys.\ Rev.\ {\bf D60} (1999)
023504 [arXiv:hep-ph/9802358];
%
J.R. Espinosa, A. Riotto and G.G. Ross, Nucl.\ Phys.\ {\bf B531}
(1998) 461-477 [arXiv:hep-ph/9804214];
%
S.F. King and A. Riotto, Phys.\ Lett.\ {\bf B442} (1998) 68-73
[arXiv:hep-ph/9806281];
%
J. Lesgourgues, Phys.\ Lett.\ {\bf B452} (1999) 15-22
[arXiv:hep-ph/9811255];
%
E. Halyo, Phys.\ Lett.\ {\bf B454} (1999) 223-227
[arXiv:hep-ph/9901302]; Phys.\ Lett.\ {\bf B461} (1999) 109-113
[arXiv:hep-ph/9905244];

\bibitem{gShybrid}
C. Panagiotakopoulos, Phys.\ Rev.\ {\bf D55} (1997) 7335-7339
[arXiv:hep-ph/9702433]; Phys.\ Lett.\ {\bf B402} (1997) 257-262
[arXiv:hep-ph/9703443];
%
A.D. Linde and A. Riotto, Phys.\ Rev.\ {\bf D56} (1997) 1841--1844
[arXiv:hep-ph/9703209];
%
L. Covi, G. Mangano, A. Masiero and G. Miele, Phys.\ Lett.\ {\bf
B424} (1998) 253-258 [arXiv:hep-ph/9707405];
%
M.~Bastero-Gil and S.~F.~ King, Phys.\ Lett.\ {\bf B423} (1998)
{27} [arXiv:astro-ph/9709502];
%
G.R. Dvali, G. Lazarides and Q. Shafi, Phys.\ Lett.\ {\bf B424}
(1998) 259-264 [arXiv:hep-ph/9710314];
%
T. Watari and T. Yanagida, Phys.\ Lett.\ {\bf B499} (2001) 297-304
[arXiv:hep-ph/0011389];

\bibitem{cmbfast}U.~Seljak and M.~ Zaldarriaga, ``A Line of Sight Approach to Cosmic
Microwave Background Anisotropies'', \apj{469}{1}{1996}; see also the site at
http://www.sns.ias.edu/~matiasz/CMBFAST/cmbfast.html.

\bibitem{verde}
L.~Verde {\it et al.}, ``First Year Wilkinson Microwave Anisotropy
Probe (WMAP) Observations: Parameter Estimation Methodology,''
arXiv:astro-ph/0302218.

\bibitem{map-params}
D.~N.~Spergel {\it et al.}, ``First Year Wilkinson Microwave
Anisotropy Probe (WMAP) Observations: Determination of
Cosmological Parameters,'' arXiv:astro-ph/0302209.

\bibitem{map-inf}
H.~V.~Peiris {\it et al.}, ``First year Wilkinson Microwave
Anisotropy Probe (WMAP) observations:  Implications for
inflation,'' arXiv:astro-ph/0302225.

\bibitem{others}
V.~Barger, H.~S.~Lee and D.~Marfatia,
``WMAP and inflation,''
arXiv:hep-ph/0302150;\\
S.~Dodelson and L.~Hui,
``A horizon ratio bound for inflationary fluctuations,''
arXiv:astro-ph/0305113;\\
W.~H.~Kinney, E.~W.~Kolb, A.~Melchiorri and A.~Riotto,
``WMAPping inflationary physics,''
arXiv:hep-ph/0305130;\\
A.~R.~Liddle and S.~M.~Leach, ``How long before the end of
inflation were observable perturbations  produced?,''
arXiv:astro-ph/0305263.

\bibitem{etaproblem}
E.J. Copeland, A.R. Liddle, D.H. Lyth, E.D. Stewart and D. Wands,
Phys.\ Rev.\ {\bf D49} (1994) 6410 [arXiv:astro-ph/9401011];
%
For a recent discussion see, N. Arkani-Hamed, H.-C. Cheng, P.
Creminelli and L. Randall, [arXiv:hep-th/0302034].

\bibitem{sugra}
E. Cremmer, {\it et. al.}, Nucl.\ Phys.\ {\bf B147} (1979) 105;
%
E. Witten and J. Bagger, Phys.\ Lett.\ {\bf B115} (1982) 202.

\bibitem{noscale}
J.R. Ellis, K. Enqvist and D.V. Nanopoulos, Phys.\ Lett.\ {\bf
B147} (1984) 99;
%
J.R. Ellis, C. Kounnas and D.V. Nanopoulos, Nucl.\ Phys.\ {\bf B
241} (1984) 406.


\end{thebibliography}
\end{document}